\documentclass{elsart}
\usepackage{natbib}
\usepackage{graphicx}
\usepackage{longtable}
\usepackage{amssymb}
\begin{document}
\begin{frontmatter}

\title{Inequalities of wealth distribution in a conservative economy}

\author[IF]{S. Pianegonda\corauthref{cor}},
\corauth[cor]{Corresponding author}
\ead{salete@if.ufrgs.br}
\author[IF]{J. R. Iglesias},
\address[IF]{Instituto de F\'{\i}sica, U.F.R.G.S., C.P. 15051, 91501-970
 Porto Alegre - RS, Brazil.}

\thanks{We acknowledge fruitful discussions with G. Abramson,
R. Donangelo, S. Gon\c{c}alves, S. Risau-Gusman and J.L. Vega, and
financial support from CNPq, Brazil}

\begin{abstract}
We analyze a conservative market model for the competition among
economic agents in a close society. A minimum dynamics ensures
that the poorest agent has a chance to improve its economic
welfare. After a transient, the system self-organizes into a
critical state where the wealth distribution have a minimum
threshold, with almost no agent below this poverty line, also,
very few extremely rich agents are stable in time. Above the
poverty line the distribution follows an exponential behavior. The
local solution exhibits a low Gini index, while the mean field
solution of the model generates a wealth distribution similar to
welfare states like Sweden.
\end{abstract}


\end{frontmatter}

\section{Introduction}
\label{} The study of the distribution of the income of workers,
companies and countries started a little more than a century ago
when Italian economist Vilfredo Pareto investigated data of
personal income for different European countries and found a power
law distribution that seems not to depend on different economic
conditions of the countries. In 1887, in his book {\it Cours
d'Economie Politique} \cite{Pareto} he asserted that in all
countries and times the distribution of income and wealth follows
a power law behavior where the cumulative probability $P(w)$ of
people whose income is at least $w$ is given by $P(w) \propto
w^{-\alpha}$, where the exponent $\alpha$ is named Pareto index.
This universal power law is now called Pareto law and he found
himself that the index assumes values typically close to 1.5 for
several countries. The reduction of the slope $\alpha$ denotes a
higher inequality of the income distribution.

However, recent data suggest that Pareto's distribution does not
agree with observed data over the middle range of income, but it
is generally accepted that it provides a good fit to the
distribution of high range of income. Many other distributions of
income were proposed, supported by international empirical data,
and the distribution behaves in a log-normal, exponential or Levy
way \cite{Bouchaud} in the middle income range. For instance, there
are data from Japan and USA that are fitted by a log-normal
distribution  in middle range plus a power law for high income
one \cite{Souma2002}, data from United Kingdom income and wealth
fitted by an exponential law (middle range) and a power law (high
range) \cite{Dragulescu} and data from Brazil \cite{Brasil}
worker's income and enterprise capital that are fitted by a
log-normal plus a power law for high range of
income \cite{Brasil2}.

One of the goals of the study of wealth distribution is to
determine the social inequalities implied in a given model of
economic exchanges. Many ways to quantitatively measure the
fairness of a given income distribution have been proposed in the
economic science. Among them, the Gini index is well known and
frequently used. In brief, for any given wealth density function
$p(x)$, with mean $\rho$ and a given number $\nu \in [0,1]$,
$P(\leq \nu)=\int_{0}^{\nu} p(x)dx$ represents the cumulative
fraction of population with wealth bellow $\nu$ and Lorenz
function $L(\nu)$ is defined as the fraction of the total wealth
which is attributed to the poorest $100\nu$ percentage of the
society. It easy to show that
$L(\nu)=\frac{1}{\rho}\int_{0}^{\nu}xp(x)dx$. Gini index can be
then defined as $G=2\int_{0}^{1}(\nu-L(\nu))d\nu$ \cite{Gini}. It
assumes values between 0 percent (perfect equality) and 100
percent (perfect inequality).

Recently we developed a simple parameter free model for the
competition among different economic agents within a conservative
model. Results for the wealth distribution has been published
\cite{Pianegonda} and also considering the effect of a small world
network \cite{Iglesias}. Within this model we obtained an
exponential Gibbs-style distribution for the wealth, with a finite
lower threshold or poverty line, that is characteristic of models
with extremal dynamics. In spite of the fact that this kind of
distribution seems to be at stake when compared with Pareto law,
the idea of a society that take measures in order to improve the
situation of the most impoverished is compatible with the
propositions of John Rawls, in his book ``A Theory of Justice''
\cite{Rawls}, directed towards an inventive way of securing
equality of opportunity as one of the basic principles of justice.
He asserts that {\it no redistribution of resources within...a state
can occur unless it benefits the least well-off} and this should
be the only way to prevent the stronger (or richer) from
overpowering the weaker (or poorer). It seems that a minimum
dynamics, as that proposed in the Bak-Sneppen model \cite{BAK},
when applied to economics, is a way to ensure to the poorest agent
a chance to improve its situation.

In the next section we resume the principal assumptions of the
model and the principal results of the numerical simulations. The
calculation of the Gini index and the comparison with wealth
distribution of several countries will be presented in section
\ref{ineq} along with the conclusions

\section{The Conservative Exchange Market Model}
\label{def} The Conservative Exchange Market Model (CEMM)
\cite{Pianegonda} is a simple macroeconomic model that consists of
an one-dimensional lattice with $N$ sites and periodic boundary
conditions, where every site represents an economic agent
(individuals, industries or countries). To each agent it is
assigned some wealth-parameter that represents its welfare, like
the GNI for countries or salary for individuals. We choose an
initial configuration where the wealth is a number between 0 and 1
distributed randomly among agents. The dynamics of the system is
supported on the idea that some measure should be taken to modify
the situation of the poorest agent. In this context, we model this
process by a minimum dynamics \cite{BAK}: at each time step, the
poorest agent, i.e., the one with the minimum wealth, will take
some action, trying to improve its economic state. Since the
outcome of any such measure is uncertain, the minimum suffers a
random change in its wealth \cite{Pianegonda}. We assume that
whatever wealth is gained (or lost) by the poorest agent it will
be at the expense of its neighbors and we assume that it is
equally debited (or credited) between its two nearest neighbors on
the lattice, making the total wealth constant. So, the CEMM
dynamic can be reduced as follows:
\begin{itemize}
\item Find out the site with minimum wealth $x_{min}$; \item
Substitute $w(x_{min})$ to $w_{new}$, a new random number between
0 and 1 extracted from an uniform distribution; \item Make the
difference, $\Delta=w_{new}(x_{min})-w(x_{min})$ \item Updating
the neighbors: \\$w(x_{min}-1)=w(x_{min}-1)-\Delta/2$\\
$w(x_{min}+1)=w(x_{min}+1)-\Delta/2$\\ Note that if $\Delta$ is
positive (negative) the wealth of the neighbors will be reduced
(increased) by $\Delta/2$.
\end{itemize}

\begin{figure}
\begin{center}
\includegraphics*[scale=0.31]{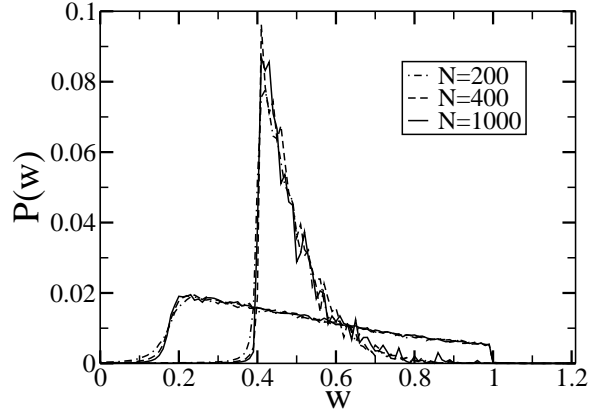}
\end{center}
\caption{Distribution wealth $P(w)$ for the model with
interactions to nearest neighbors (upper curves) and with global
interactions (lower curves). In both cases we have superposed
scaled data corresponding to\textsl{} $N=200$, $400$, $1000$.
The histograms were built using $10^5$ consecutive
states, after a transient of $10^6$ steps has elapsed.}
\label{distrisal}
\end{figure}

Numerical simulations on this model showed that, after a
relatively long transient, the system arrives at a self organized
critical (SOC) state with a stationary wealth distribution (figure
\ref{distrisal}\cite{Pianegonda}) in which almost all agents are
beyond a certain threshold. On Fig\ref{distrisal} we represent
both the cases of local exchanges and of global exchanges (mean
field solution); in the former case the poverty line is $\eta_T
\approx 0.4$ and above this threshold the distribution of agents
is an exponential $P(w) \approx exp(-w^{2}/2 \sigma^2)$, with
$\sigma =0.228$ \cite{Iglesias}, i.e., there are exponentially few
rich agents while the mass of them remains in what we call a {\it
middle class}. In the second case the figure show the globally
coupled ({\it mean field}) exchanges (lower curves). This
corresponds to a situation in which the agents with which the
exchange takes place are chosen at random and not based on
geographical proximity. This mean field solution exhibits a lower
threshold, $\eta_T \approx 0.2$, and beyond it also an exponential
distribution, $P(w) \approx exp(-w^{2}/2 \sigma^2)$, but with a
higher value of $\sigma =0.567$ \cite{Iglesias}.

\begin{figure}
\begin{center}
\includegraphics*[scale=0.35]{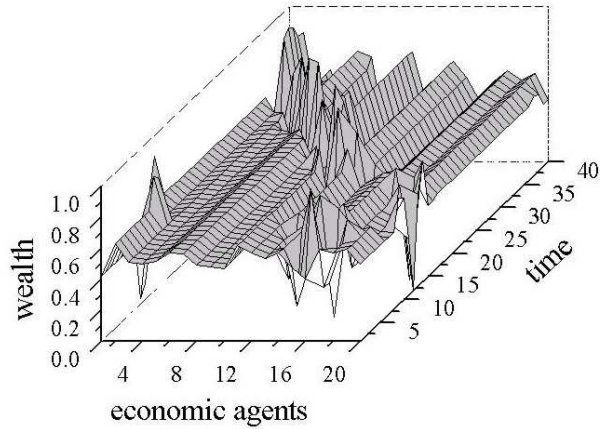}
\end{center}
\caption{Snapshot of avalanche propagation for the local case. The
sites with maximum wealth are very stable in time.} \label{g3d}
\end{figure}

The temporal evolution, in the SOC state, of the position of the
system's minimum and maximum wealth is quite different. While the
site of minimum wealth is changing continuously, generating
avalanches of all sizes among neighbors, the wealthiest site is
stable over relatively long periods of time \cite{Pianegonda} as
it is represented by the lines in the figure \ref{g3d}. Indeed,
even when affected by an avalanche the maximum can frequently
recover its status after a short time. Not only the absolute
maximum is stable, but also a privileged group, whose wealth is
around the same value of the maximum, remains in its prosperous
position for quite a while. This group is also privileged by
having high wealth values, well above the market average. In the
mean-field case the rich sites are also stable, but over shorter
periods of time.

\begin{figure}
\begin{center}
\includegraphics*[scale=0.31]{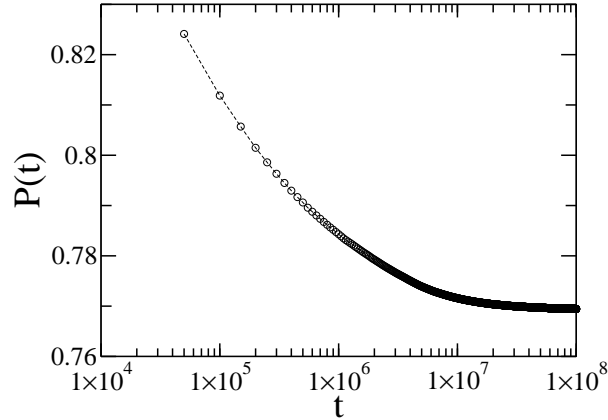}
\end{center}
\caption{Probability of an agent become wealthier in a time step
for a market with $N=1000$ agents.} \label{prob}
\end{figure}

We also observed that the probability that the minimum asset agent
improve its situation in a time step is relatively high. It is
very high at the beginning of the simulation, then it decreases
during the transient to finally converge to a finite value,
$p\approx 0.77$ as it is shown on figure \ref{prob}. We remark
that the probability is well above 0.5, so this minimum dynamic
has the effect of improve, in the average, the situation of the
poorer agents.

\section{Income inequality}
\label{ineq} Inequalities are related to social and economic
phenomena and is a subject of much debate. The fraction of rich
and poor people in a country depends on the inequalities of the
income distribution. The standard way to determine the inequality
of a wealth distribution is with the help of quintile tables that
show the percentage of national income received by equal
percentiles of individuals or households ranked by their per
capita income levels. For example in a welfare state as Sweden the
richest 20 percent of the population receives about 4 times more
than the poorest quintile, while in a {\it ruthless} capitalistic
country like Brazil the richest quintile receives more than 25
times more than the poorest quintile. Table 1 gives the quintile
data for a few countries, while full data can be obtained in the
World Bank site\cite{WB2,tabela}.

\begin{table}[h]
\centering
\begin{tabular}{lcccc} \hline
& USA (1997) & Sweden (1992) & Spain (1990) & Brazil (1998) \\ \hline
Poorest 20\% & 5.2 & 9.6 & 7.5& 2.2 \\ \hline
Fourth 20\% & 10.5 & 14.5 & 12.6 & 5.4 \\ \hline
Third 20\% & 15.6 & 18.1 & 17.0 & 10.1 \\ \hline
Second 20\% & 22.4 & 23.2 & 22.6 & 18.3 \\ \hline
Richest 20\% & 46.4 & 34.5 & 40.3 & 64.1 \\ \hline
Gini index & 40.8 & 25.0 & 32.5 & 60.7 \\ \hline
\label{table1}
\end{tabular}
\centering \caption {Percentage of income in each population
quintile for some selected countries \cite{tabela}.}
\end{table}

\begin{figure}
\begin{center}
\includegraphics*[scale=0.37]{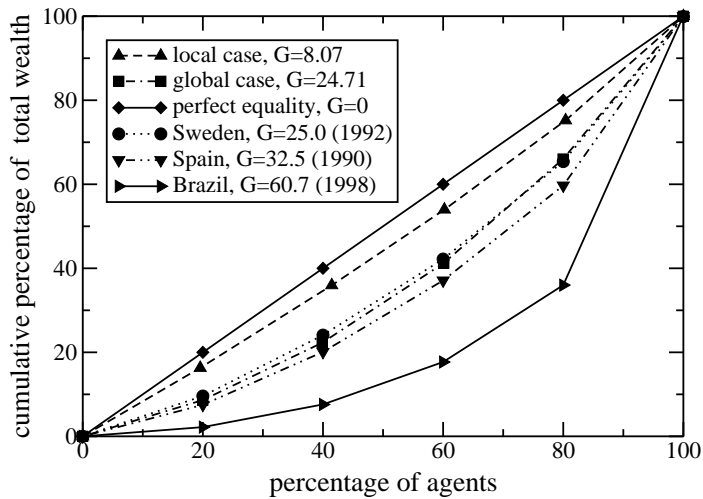}
\end{center}
\caption{The Lorenz curves and the Gini index for ideal, local and
the global case. Also represented are the World Bank data for
Brazil, Sweden and Spain \cite{tabela}.} \label{gini}
\end{figure}

To measure income inequality in a country and to make a comparison
among countries more accurately, economists use Lorenz curves and
Gini indexes. A Lorenz curve plots the cumulative percentages of
total income against the cumulative percentages of recipients. In
Fig \ref{gini} we have represented the Lorenz curves for Brazil,
Sweden and Spain. For comparison, we have plotted in same figure
the Lorenz curve of an hypothetical absolutely equalitarian income
distribution that is the straight line with slope equal 1. This
figure shows that Brazil's Lorenz curve strongly deviates from the
hypothetical line of absolute equality while that for Sweden lies
much nearer and that for Spain in the middle of both. This means
that of these three countries, Brazil has the highest income
inequality and Sweden the lowest. We also constructed Lorenz
curves for the local and global cases of CEMM model (figure
\ref{gini}). Clearly we see that the global case have a Lorenz
curve very similar to Sweden, while the local case is close to
perfect equality.

The Gini index is even more convenient than a Lorenz curve when
the task is to compare income inequality among many countries; it
is calculated as the area between a Lorenz curve and the
hypothetical line of absolute equality. It assumes values between
0 percent (perfect equality) and 100 percent. Of course, real Gini
indexes are always in between. We calculated the Gini index for
the local and global cases. For the local case, the Gini index is
8.07, close to perfect equality, while its value is 24.71 for the
global one (figure \ref{gini}). This last value is very close to
that Sweden (25.0), and also close to values of other European
developed countries.

Concluding, the analysis of the Gini index shows that
globalization (mean field case) increases the economic
inequalities in a conservative economy. Moreover, the poverty line
in global case is well bellow that of the local case (figure
\ref{distrisal}). Another interesting aspect is that the minimum
dynamic favors the poorest agent, so promoting wealth
redistribution in the sense defined by Rawls \cite{Rawls}. Also,
from a strictly economic point of view, high inequality means
lower exchanges, and so it implies in a reduction of the economic
activity.

\end{document}